\documentclass[twocolumn]{article} 

\usepackage{PRIMEarxiv}
\usepackage[utf8]{inputenc}
\usepackage[T1]{fontenc}
\usepackage{hyperref}
\usepackage{url}
\usepackage{amsfonts}
\usepackage{nicefrac}
\usepackage{microtype}
\usepackage{fancyhdr}
\usepackage{graphicx}

\usepackage{amsmath,amssymb}
\usepackage{algorithmic}
\usepackage{textcomp}
\usepackage{xcolor}
\usepackage{bm}
\usepackage{booktabs}
\usepackage{siunitx}
\usepackage{multirow}
\usepackage{adjustbox}
\usepackage{comment}

\graphicspath{{media/}}    

\usepackage{geometry}
\title{Scheduling Analysis of UAV Flight Control Workloads on PREEMPT\_RT Linux Using a Raspberry Pi 5}

\author{
  Luiz Giacomossi \\
  Mälardalen University \\
  Västerås, Sweden \\
  \texttt{luiz.giacomossi@mdu.se}
  \and
  Håkan Forsberg \\
  Mälardalen University \\
  Västerås, Sweden \\
  \texttt{hakan.forsberg@mdu.se}
  \and
  Baran Çürüklü \\
  Mälardalen University \\
  Västerås, Sweden \\
  \texttt{baran.curuklu@mdu.se}
  \and
  Ivan Tomasic \\
  Mälardalen University \\
  Västerås, Sweden \\
  \texttt{ivan.tomasic@mdu.se}
  \and
  Tommaso Cucinotta \\
  ReTiS Lab, Scuola Superiore Sant'Anna \\
  Pisa, Italy \\
  \texttt{tommaso.cucinotta@santannapisa.it}
}

\begin{document}

\twocolumn[
  \begin{@twocolumnfalse}
    \maketitle
    \begin{abstract}
      Modern UAV architectures increasingly aim to unify high-level autonomy and low-level flight control on a single General-Purpose Operating System (GPOS). However, complex multi-core System-on-Chips (SoCs) introduce significant timing indeterminism due to shared resource contention. This paper performs an architectural analysis of the \texttt{PREEMPT\_RT} Linux kernel on a Raspberry Pi 5, specifically isolating the impact of kernel activation paths—deferred execution (SoftIRQs) versus real-time direct activation—on a 250\,Hz control loop. Results show that under heavy stress, the standard kernel is unsuitable, exhibiting worst-case latencies exceeding 9\,ms. In contrast, \texttt{PREEMPT\_RT} reduced the worst-case latency by nearly 88\% to under 225\,$\mu$s, enforcing a direct wake-up path that mitigates OS noise. These findings demonstrate that while \texttt{PREEMPT\_RT} resolves scheduling variance, the residual jitter on modern SoCs is primarily driven by hardware memory contention.
    \end{abstract}
    
    
    \vspace{1cm} 
  \end{@twocolumnfalse}
]

\section{Introduction}
Unmanned Aerial Vehicles (UAVs) are evolving from remote-controlled aircraft into autonomous edge-computing nodes capable of complex onboard tasks including SLAM, computer vision, and swarm coordination \cite{survey_uavs, CETDS_IEEE, Search_IAI}. Historically, these systems relied on a dual-processor architecture: a high-performance "companion computer" for autonomy and a reliable low-power microcontroller (MCU) for flight stabilization. However, constraints on size, weight, and power (SWaP) favor \textit{unified architectures}, where a single high-performance System-on-Chip (SoC) handles both mission logic and the safety-critical Flight Control System (FCS) \cite{kangunde2021review}.

The challenge lies in the operating system. While the complexity of autonomy stacks necessitates a General-Purpose Operating System (GPOS) like Linux, standard kernels are optimized for throughput, not the determinism required by 250+ Hz control loops \cite{linux_raspberry}. Unlike MCUs, modern high-performance SoCs (e.g., those based on the ARM Cortex-A76) introduce non-trivial sources of indeterminism, including deep cache hierarchies, out-of-order execution, and shared memory controllers. On such platforms, unpredictable scheduling latencies and kernel housekeeping tasks can lead to control loop jitter that threatens flight stability \cite{control_design_reference, 9999639}.

This trade-off motivates the \texttt{PREEMPT\_RT} patch-set, which transforms the Linux kernel into a hard real-time system by converting spin-locks into rt-mutexes and forcing interrupt handlers to run as preemptible kernel threads \cite{preempt_survey}. While the theoretical benefits of \texttt{PREEMPT\_RT} are well-established, existing literature \cite{linux_raspberry} predominantly focuses on older, simpler single-board computers (e.g., Raspberry Pi 3). These studies tend to overlook the behavior of modern heterogeneous multi-core architectures where shared resource contention (L3 cache, DRAM bandwidth) and complex interrupt routing significantly impact real-time guarantees.

This paper bridges that gap by providing an architectural schedulability analysis of flight control workloads on the Raspberry Pi 5. We select this platform not merely for its popularity, but because its BCM2712 SoC (quad-core Cortex-A76) exhibits the Out-of-Order execution and memory hierarchy complexity typical of modern edge-AI platforms. We isolate the CPU scheduler from hardware I/O to answer two fundamental questions:
\begin{enumerate}
    \item \textbf{Architectural Mechanism:} How do the distinct task activation paths (Deferred SoftIRQ vs. Direct Threaded IRQ) in standard and real-time kernels dictate worst-case latency?
    \item \textbf{Resource Contention:} To what extent does the \texttt{PREEMPT\_RT} kernel mitigate jitter caused by shared cache and memory bandwidth contention under heavy autonomy-like loads?
\end{enumerate}

The main contributions of this work are: 
1) A quantitative performance benchmark of a 250\,Hz control loop on a modern multi-core SoC, isolating scheduler behavior from bus latency; 
2) An architectural analysis identifying that the standard kernel's reliance on Deferred Execution (SoftIRQs) is the primary cause of $>9\,ms$ latency spikes, whereas \texttt{PREEMPT\_RT}'s Direct Activation path reduces this by 88\% (from 1.84\,ms to $< 225\,\mu$s); 
3) Empirical validation that while \texttt{PREEMPT\_RT} effectively bounds scheduling latency, residual jitter on modern SoCs is driven by hardware memory contention, providing a computational baseline for unified UAV architectures.

\section{Background and Related Work}
\label{sec:Background}
This section reviews the fundamentals of real-time UAV control, examines existing software architectures, and contextualizes this work.

\subsection{Overview of Real-Time UAV Control Systems}
Flight control systems are hard real-time systems. They execute nested control loops, with high-frequency inner loops (200--500\,Hz) for attitude stabilization and slower outer loops (50--100\,Hz) for navigation \cite{giernacki2017crazyflie, bouabdallah2007design, beard2012small}. The stability of these loops depends on the timing guarantees provided by the operating system, captured by the principle of \textit{determinism}. In real-time systems, determinism means that task execution times are bounded so that deadlines can be guaranteed under specified conditions. This relies on two key properties:

\begin{itemize}
    \item \textbf{Bounded Latency:} Scheduling latency is the delay between a task becoming ready and its dispatch to the CPU. Large or unpredictable latencies reduce the control phase margin and can destabilize the system \cite{9999639}. For a 250\,Hz loop (4\,ms period), latencies exceeding a fraction of the period (e.g., $>500\,\mu$s) can destabilize the control loop.
    
    \item \textbf{Low Jitter:} Jitter is the variation in latency across executions. Excessive jitter makes the control loop period inconsistent, complicates controller tuning (e.g., needs lower gains), and creates unpredictable dynamics \cite{smeds2012effect}.
\end{itemize}

\subsection{Architectures for Embedded FCS}
Embedded FCS architectures have evolved to support the increasing demand for onboard intelligence while managing SWaP (Size, Weight, and Power) constraints:

\begin{itemize}
    \item \textbf{Microcontroller + RTOS:} The standard combines a microcontroller (e.g., STM32) with a Real-Time Operating System (RTOS) like FreeRTOS or NuttX. This provides excellent determinism (jitter typically $<10\,\mu$s) but lacks the memory and compute throughput required for onboard computer vision or SLAM \cite{PX4_Arch}. 
    
    \item \textbf{Dual-Processor Setup:} A high-level companion computer (e.g., Raspberry Pi, Jetson) handles perception and communicates with a separate low-level RTOS microcontroller via a serial link (e.g., MAVLink) \cite{nvidia_pixhawk}. While robust, this adds weight, wiring complexity, and inter-process communication (IPC) latency.

    \item \textbf{Unified GPOS-Based System:} Control and high-level autonomy are integrated onto a single high-performance SoC running a GPOS, typically Linux with real-time extensions \cite{martensson2016flying, RT_Helicopter}. This reduces hardware complexity but introduces the challenge of ensuring real-time determinism on complex multi-core architectures where cache and memory bus contention can cause unpredictable delays. This architecture is the focus of our investigation.    
\end{itemize}

\subsection{Linux Scheduling Policies}
The Linux kernel scheduler implements several policies. The default policy, \texttt{SCHED\_OTHER} (based on the Completely Fair Scheduler, CFS), is optimized for throughput and fair time-slicing but offers no timing guarantees \cite{love2010linux}. For real-time applications, Linux provides alternative policies summarized in Tab.~\ref{tab:sched_policies}. Foundational works \cite{preempt_survey} have established that standard kernels suffer from priority inversion and non-preemptible sections (kernel locks), necessitating the \texttt{PREEMPT\_RT} patch to enforce bounded latencies.

\begin{table}[ht]
    \centering
    \caption{Linux Kernel Scheduling Policies and Characteristics.}
    \label{tab:sched_policies}
    \resizebox{\columnwidth}{!}
    {%
    \begin{tabular}{|l|l|l|l|}
        \hline
        \textbf{Policy} & \textbf{Type} & \textbf{Priority Mechanism} & \textbf{Key Feature} \\
        \hline
        \hline
        \texttt{SCHED\_OTHER} & Time-Sharing & Dynamic (`nice` value) & Fair CPU sharing for throughput \\
        \texttt{SCHED\_FIFO} & Real-Time & Static (1-99) & Runs until block/yield \\
        \texttt{SCHED\_RR} & Real-Time & Static (1-99) & Round-robin time-slicing \\
        \texttt{SCHED\_DEADLINE} & Real-Time & Dynamic (EDF) & Temporal isolation via reservation \\
        \hline
    \end{tabular}%
    }
\end{table}

\subsection{Related Work and Gap Analysis}
Research into unified Linux-based flight control generally follows three tracks: dual-kernel extensions, virtualization, and native preemption.

\subsubsection{Dual-Kernel (Xenomai) vs. Native Preemption}
To overcome Linux's non-determinism, approaches like \textbf{Xenomai (Cobalt)} use a dual-kernel architecture, running a real-time microkernel alongside Linux. While Xenomai typically offers lower latency than \texttt{PREEMPT\_RT} by bypassing the Linux scheduler entirely \cite{brown2010fast}, it requires custom drivers (RTDM) and creates a complex split-system development environment. 
In the context of UAVs, we prioritize \texttt{PREEMPT\_RT} because it maintains the standard Linux programming model. This allows standard robotics frameworks (e.g., ROS 2, MAVROS) and device drivers (e.g., V4L2 for cameras) to be used without modification, simplifying unified autonomy development.

\subsubsection{Isolation via Virtualization}
Yang and Shinjo \cite{yang2020obtaining} proposed a compounded RTOS (cRTOS) using \textbf{Jailhouse} to partition hardware between Linux and NuttX. While this ensures isolation, it effectively mimics the dual-processor constraint in software, limiting the flexibility of resource sharing between the autonomy and control layers.

\subsubsection{Hardware Complexity Gap}
Most \texttt{PREEMPT\_RT} benchmarks focus on older, simpler Single Board Computers (e.g., RPi 3, Cortex-A53), establishing a baseline of approx.~150\,$\mu$s worst-case latency under moderate load \cite{adam2021real}. However, these studies do not account for modern \textit{Out-of-Order (OoO)} architectures like the Raspberry Pi 5's Cortex-A76. On these platforms, despite higher clock speeds, shared resource contention (L3 Cache/DRAM) becomes a primary source of timing violation \cite{iorga2020slow}. Our work builds on this by quantifying how the \texttt{PREEMPT\_RT} scheduler mitigates interference in these complex, heterogeneous environments.

\section{Flight Control Architecture}
\label{sec:Architecture}

\label{subsec:flight_control}

\begin{figure}[htpb]
    \centering
    \includegraphics[width=0.95\columnwidth]{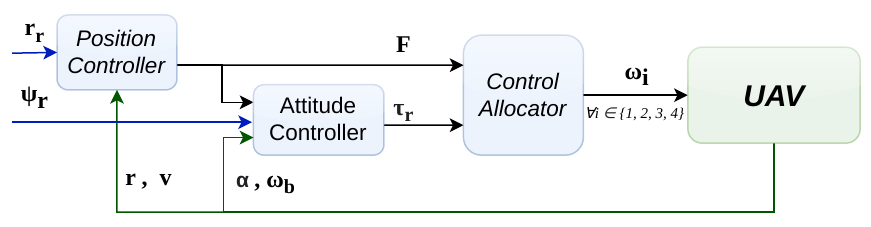}
    \vspace{-3mm}
    \caption{Architecture of the UAV Flight Control System. The \textit{Position Controller} generates the desired force vector $\mathbf{F}$ based on position and velocity errors. The \textit{Attitude Controller} computes the desired torque vector $\boldsymbol{\tau}_r$ from attitude and angular velocity errors. The \textit{Control Allocator} maps these commands to individual motor speeds $\omega_i$ for $i \in \{1,2,3,4\}$.}

    \label{fig:fcs_overview}
\end{figure}

The FCS of the quadcopter is implemented using the cascaded architecture, as described in~\cite{optimization_fcs} and depicted in Fig.~\ref{fig:fcs_overview}. The outer loop features a \textit{Position Controller}, which takes the position reference $\mathbf{r}_r$, the current position $\mathbf{r}$, and the current linear velocity $\mathbf{v}$ as inputs. It calculates the total desired force vector $\mathbf{F}$, required to track the position reference.

The inner loop consists of the \textit{Attitude Controller}. This controller receives the desired yaw angle $\psi_r$, the current attitude $\boldsymbol{\alpha}$ (i.e., Euler angles), and the current body-frame angular velocity $\boldsymbol{\omega}_b$. It uses this information to compute the desired control torque vector, $\boldsymbol{\tau}_r$. The \textit{Control Allocator} receives the total commanded force $\mathbf{F}$ and the desired torque $\boldsymbol{\tau}_r$. Its role is to decompose these high-level commands into actuator-level inputs by computing the individual rotor speeds $\omega_i$ needed to achieve the desired response within actuator limits.

Finally, these actuator commands are applied to the UAV model. In this work, to isolate the performance of the CPU scheduler from hardware I/O, the vehicle's dynamics are handled by a \textit{UAV Simulation Block}. This block simulates the vehicle's dynamics and outputs the state variables---position $\mathbf{r}$, linear velocity $\mathbf{v}$, attitude $\boldsymbol{\alpha}$, and angular velocity $\boldsymbol{\omega}_b$---which are used as feedback signals to close the control loops. For more details on the dynamic equations and control laws behind this architecture, see~\cite{optimization_fcs, bouabdallah2007design}.




\section{System Model and Methodology}
\label{sec:Implementation}
To empirically evaluate the architectural determinism of the \texttt{PREEMPT\_RT} kernel, we designed a software framework\footnote{The implementation is available at: \url{https://github.com/luizgiacomossi/Real-Time-Flight-Control-With-Linux}} that explicitly isolates scheduler activation latencies from peripheral I/O delays. This section details the platform architecture, the task model, and the stress framework used for evaluation.


\subsection{Platform Architecture: Raspberry Pi 5}
\label{subsec:rpi5_considerations}
We selected the Raspberry Pi 5 (8GB RAM) as the target platform. Unlike previous generations of Single-Board Computers (SBCs) evaluated in literature \cite{adam2021real}, the RPi 5 utilizes the Broadcom BCM2712 SoC featuring a quad-core ARM Cortex-A76 cluster. To preclude thermal throttling during stress testing, active cooling was employed, and the CPU frequency was strictly locked to 2.4\,GHz using the \texttt{performance} governor.

This choice is architecturally significant: the Cortex-A76 employs an out-of-order execution pipeline with a three-level cache hierarchy (64KB L1, 512KB L2 per core, and a shared 2MB L3). This complexity introduces sources of non-determinism—specifically shared L3 cache contention and DRAM bandwidth saturation—that are not present in simpler in-order microcontrollers. The system runs Ubuntu 24.04 LTS with two kernel configurations: the standard mainline kernel (\texttt{6.8.0-raspi}) and the fully preemptive real-time kernel (\texttt{6.8.0-raspi-realtime}).

\subsection{Real-Time Software Architecture}
The Flight Control System (FCS) is implemented as a set of concurrent, prioritized threads mapped to specific CPU cores. We focus our formal analysis on the critical high-frequency attitude control loop, as it represents the system's worst-case timing constraint. We define this periodic real-time task $\tau_{att}$ as:
\begin{equation}
    \tau_{att} = \{ C_{att}, T_{att}, D_{att}, P_{att} \}
\end{equation}
Where execution time $C_{att}$ varies based on load, period $T_{att} = 4000\,\mu s$ (250\,Hz), deadline $D_{att} = T_{att}$, and priority $P_{att}$ is a variable determined by the experimental configuration (see Section \ref{sec:Experimental}). To mitigate sources of OS-induced jitter, we apply an isolation strategy:
\begin{itemize}
    \item \textbf{CPU Isolation:} We utilize the boot parameter \texttt{isolcpus=2,3} to exclude Cores 2 and 3 from the SMP balancing algorithms.
    \item \textbf{Task Pinning:} The critical task $\tau_{att}$ is pinned exclusively to Core 2 via `pthread\_setaffinity\_np`.
    \item \textbf{Inter-Thread Isolation:} Lower-criticality tasks (logging, telemetry, and the position controller $\tau_{pos}$) are pinned to Core 3, ensuring they do not preempt the attitude loop.
\end{itemize}
The wake-up mechanism relies on \texttt{clock\_nanosleep()} using \texttt{CLOCK\_MONOTONIC}, which provides high-resolution timing backed by the hardware \texttt{hrtimer}.

\subsection{Workload Injection Framework}
\label{subsec:workload}
To analyze the architectural limits of the scheduler under worst-case conditions, we utilized \texttt{stress-ng}\footnote{The \texttt{stress-ng:}\url{github.com/ColinIanKing/stress-ng}}tool to synthesize a unified interference profile. This profile concurrently executes distinct stressor components (Tab.~\ref{tab:stress_profiles}).


\begin{table}[ht]
    \centering
    \caption{Components of the Stress Profile used for Evaluation.}
    \label{tab:stress_profiles}
    \resizebox{\columnwidth}{!}
    {%
    \begin{tabular}{|l|l|l|}
        \hline
        \textbf{Component} & \textbf{Stressor Mechanism} & \textbf{Targeted Bottleneck} \\
        \hline
        \hline
        \textbf{Compute} & \texttt{cpu-matrix} (4 workers) & ALU pipeline saturation \\
        \hline
        \textbf{Memory} & \texttt{vm} (75\% RAM, 2 workers) & \textit{L3 Cache Thrashing} \& DRAM Bandwidth \\
        \hline
        \textbf{Kernel} & \texttt{switch, fork} (High rate) & Runqueue lock contention \& \textit{SoftIRQ Latency} \\
        \hline
    \end{tabular}%
    }
\end{table}

The unified profile has three components: Compute, which tests the scheduler's ability to preempt compute-bound threads; Memory, which triggers L3 cache evictions to simulate interference from memory-heavy autonomy workloads (e.g., Computer Vision) on neighboring cores; and Kernel, which floods the system with context switches and interrupts to evaluate the kernel's handling of the "Deferred Activation Path" (SoftIRQs) described in Section \ref{sec:Results}.

\subsection{Methodological Constraints}
\label{subsec:data_acquisition_latency_components}
A challenge in any real-world FCS is the management of I/O latency (e.g., SPI bus transfers for IMU data). While hardware transfers are deterministic, driver stacks introduce variable delays. However, scheduler determinism is the necessary antecedent condition for real-time control; if the kernel cannot dispatch the control task within the deadline, the speed of the I/O driver is irrelevant.
Therefore, this study intentionally decouples scheduler latency from I/O. For the experiments, $\tau_{att}$ executes the full control law mathematics but bypasses physical SPI transactions. This ensures that the measured jitter is attributable solely to CPU scheduling and kernel activation paths, enabling a precise architectural comparison of \texttt{PREEMPT\_RT} versus standard Linux.

\section{Experimental Evaluation}
\label{sec:Experimental}

To quantify the performance difference between the standard Linux kernel and its \texttt{PREEMPT\_RT} counterpart, we designed an experiment to measure the scheduling latency and jitter of a representative high-frequency UAV control loop. The inner attitude stabilization loop of our FCS was set to 250\,Hz (4\,ms period), a common and demanding frequency for agile multi-rotor platforms that is fast enough to require real-time determinism but slow enough to be achievable on embedded hardware if the operating system is sufficiently responsive~\cite{bouabdallah2007design, giernacki2017crazyflie}. For all tests, this critical task was pinned to an isolated CPU core (core 2) on the Raspberry Pi 5, with memory locking enabled to prevent swapping delays.

Crucially, the experimental matrix detailed in Tab.~\ref{tab:experiments} was executed twice: once on a standard Linux kernel to establish a baseline, and once on a kernel with \texttt{PREEMPT\_RT} to evaluate its real-time capabilities. This yields a total of 32 unique test runs. The 16 configurations tested on each kernel were:
\begin{itemize}
    \item \texttt{SCHED\_OTHER} (at nice levels 0 and -19) to establish a non-real-time baseline.
    \item \texttt{SCHED\_FIFO} and \texttt{SCHED\_RR} (at static priorities 50 and 99) to assess POSIX real-time policies.
    \item \texttt{SCHED\_DEADLINE} with runtimes of 400\,$\mu$s and 800\,$\mu$s, chosen to represent a tight and a generous budget based on measured task execution time. The period and deadline were set to 4\,ms to match the 250\,Hz control loop.
\end{itemize}

\begin{table}[ht]
\centering
\caption{Experimental configurations executed on \textbf{each} kernel, for a total of 32 tests. For \texttt{SCHED\_DEADLINE}, the deadline and period set to 4.0\,ms, matching the control loop's period.}
\label{tab:experiments}
\resizebox{\columnwidth}{!}{%
\begin{tabular}{|l|l|c|c|c|}
\hline
\textbf{Scheduler} & \textbf{Parameter Values} & \textbf{Period (ms)} & \textbf{Stress-ng} & \textbf{Tests per Kernel} \\
\hline
\texttt{SCHED\_OTHER }    & Nice = \{0, -19\}                    & 4.0  & Off / On & $2 \times 2 = 4$ \\
\texttt{SCHED\_FIFO }     & Priority = \{50, 99\}                & 4.0  & Off / On & $2 \times 2 = 4$ \\
\texttt{SCHED\_RR }       & Priority = \{50, 99\}                & 4.0  & Off / On & $2 \times 2 = 4$ \\
\texttt{SCHED\_DEADLINE } & Runtime = \{400, 800\}~$\mu$s        & 4.0  & Off / On & $2 \times 2 = 4$ \\
\hline
\multicolumn{4}{|r|}{\textbf{Total}} & \textbf{16 per kernel} \\
\hline
\end{tabular}
}
\end{table}

Each configuration was tested with and without background load using the \texttt{stress-ng} tool. The tool was configured to simulate a realistic system load representative of a modern UAV autonomy stack. To this end, we launched a set of stressors, including four CPU workers executing matrix multiplications to simulate tasks like path planning, two virtual memory stressors allocating 75\% of available memory to represent computer vision workloads stressing the system like real autonomy stacks would, without destabilizing the testbed, and multiple inter-process communication (IPC) stressors to mimic the data flow in frameworks like ROS. To ensure maximal determinism for the benchmark, the kernel's real-time runtime throttling was disabled by setting \texttt{sched\_rt\_runtime\_us} to -1, and all CPU cores were locked to their maximum frequency (2.4\,GHz) via the \texttt{performance} governor. Each of the 32 tests executed \(10,000\) iterations of the control loop, allowing for a detailed statistical analysis. 

It is important to note that, all latency metrics are presented in microseconds (\textmu s). These values were derived from high-resolution timers with nanosecond precision. The conversion to microseconds can introduce a visual quantization effect, observable as discrete steps in the time-series plots.


\subsection{Analysis of Scheduling Determinism}
\label{subsec:analysis-methodology}

To investigate the origins of the non-determinism observed in the benchmark, particularly for \texttt{SCHED\_OTHER}, we designed two follow-up experiments. The objective was to distinguish between interference from high-level system services and fundamental architectural behaviors of the kernel scheduler. The experiments were:
\begin{enumerate}
    \item \textbf{System-Level Isolation Analysis:} The benchmarks for the \texttt{SCHED\_OTHER} policy were re-executed after disabling the graphical user interface (GUI) by switching the system to the \texttt{multi-user.target} runlevel. This isolates the scheduler's performance from a primary source of non-deterministic, high-level system load.
    
    \item \textbf{Task Activation Path Analysis:} A low-level kernel trace analysis was performed using the \texttt{perf} tool. This analysis was designed to observe and compare the task wakeup latency and the sequence of kernel events for a real-time policy (\texttt{SCHED\_FIFO}) versus a standard policy (\texttt{SCHED\_OTHER}).
\end{enumerate}

\section{Results}
\label{sec:Results}

Tab.~\ref{tab:scheduling_performance} presents the statistical results of the 32 experiments. While \texttt{PREEMPT\_RT} paired with \texttt{SCHED\_OTHER} is included for completeness, its fundamental lack of real-time guarantees excludes it from further plots and discussion. The remaining real-time findings are visualized via latency distributions (Fig.~\ref{fig:rt_schedulers_boxplot}) and detailed time-series traces (Figs.~\ref{fig:sched_other_latency}-\ref{fig:sched_deadline_latency_rt}).

\begin{table}[htbp]
\centering
\caption{Unified Performance Comparison of Standard vs. \texttt{PREEMPT\_RT} Kernels Under a 250\,Hz Control Loop Task. All metrics are in microseconds (µs). Note the reduction in worst-case (Max) latency for all real-time policies when using the \texttt{PREEMPT\_RT} kernel under system stress.}
\label{tab:scheduling_performance}
\resizebox{\columnwidth}{!}{%
\begin{tabular}{|c|l|l|c|r|r|r|r|r|r|}
\hline
\textbf{Scheduler} & \textbf{Parameters} & \textbf{Kernel} & \textbf{Stress} & \textbf{Mean (µs)} & \textbf{Median (µs)} & \textbf{Max (µs)} & \textbf{StdDev (µs)} & \textbf{P90 (µs)} & \textbf{P99 (µs)} \\
\hline
\hline
\textbf{OTHER} & Nice 0 & Standard & No & 58.53 & 54.00 & 2724.00 & 73.89 & 57.00 & 72.00 \\
 & & Standard & \textbf{Yes} & \textbf{274.01} & \textbf{103.00} & \textbf{8626.00} & \textbf{563.32} & \textbf{451.00} & \textbf{3070.00} \\
\cline{2-10}
 & & PREEMPT\_RT & No & 59.64 & 58.00 & 2570.00 & 36.78 & 59.00 & 66.00 \\
 & & PREEMPT\_RT & \textbf{Yes} & \textbf{196.84} & \textbf{97.00} & \textbf{9015.00} & \textbf{414.98} & \textbf{290.00} & \textbf{2420.00} \\
\hline
\textbf{OTHER} & Nice -19 & Standard & No & 56.59 & 54.00 & 2451.00 & 45.36 & 57.00 & 70.00 \\
 & & Standard & \textbf{Yes} & \textbf{136.54} & \textbf{101.00} & \textbf{9424.00} & \textbf{227.57} & \textbf{188.00} & \textbf{642.00} \\
\cline{2-10}
 & & PREEMPT\_RT & No & 64.79 & 58.00 & 2779.00 & 88.06 & 60.00 & 90.00 \\
 & & PREEMPT\_RT & \textbf{Yes} & \textbf{110.81} & \textbf{85.00} & \textbf{4351.00} & \textbf{186.65} & \textbf{124.00} & \textbf{669.00} \\
\hline
\hline
\textbf{FIFO} & Priority 50 & Standard & No & 4.07 & 3.00 & 45.00 & 2.90 & 6.00 & 21.00 \\
 & & Standard & \textbf{Yes} & \textbf{58.37} & \textbf{39.00} & \textbf{700.00} & \textbf{48.83} & \textbf{137.00} & \textbf{201.00} \\
\cline{2-10}
 & & PREEMPT\_RT & No & 5.29 & 4.00 & 135.00 & 4.55 & 7.00 & 21.00 \\
 & & PREEMPT\_RT & \textbf{Yes} & \textbf{36.07} & \textbf{28.00} & \textbf{160.00} & \textbf{23.78} & \textbf{68.00} & \textbf{123.00} \\
\hline
\textbf{FIFO} & Priority 99 & Standard & No & 4.25 & 4.00 & 63.00 & 1.97 & 5.00 & 12.00 \\
 & & Standard & \textbf{Yes} & \textbf{49.37} & \textbf{40.00} & \textbf{1848.00} & \textbf{40.67} & \textbf{89.00} & \textbf{156.00} \\
\cline{2-10}
 & & PREEMPT\_RT & No & 4.19 & 4.00 & 100.00 & 1.90 & 4.00 & 7.00 \\
 & & PREEMPT\_RT & \textbf{Yes} & \textbf{31.33} & \textbf{25.00} & \textbf{224.00} & \textbf{21.61} & \textbf{50.00} & \textbf{130.00} \\
\hline
\hline
\textbf{RR} & Priority 50 & Standard & No & 4.56 & 4.00 & 51.00 & 2.14 & 6.00 & 13.00 \\
 & & Standard & \textbf{Yes} & \textbf{38.01} & \textbf{30.00} & \textbf{787.00} & \textbf{28.30} & \textbf{64.00} & \textbf{125.00} \\
\cline{2-10}
 & & PREEMPT\_RT & No & 4.35 & 4.00 & 76.00 & 2.53 & 5.00 & 13.00 \\
 & & PREEMPT\_RT & \textbf{Yes} & \textbf{44.09} & \textbf{32.00} & \textbf{225.00} & \textbf{31.75} & \textbf{92.00} & \textbf{152.00} \\
\hline
\textbf{RR} & Priority 99 & Standard & No & 4.76 & 4.00 & 86.00 & 3.40 & 6.00 & 16.00 \\
 & & Standard & \textbf{Yes} & \textbf{43.18} & \textbf{36.00} & \textbf{472.00} & \textbf{26.12} & \textbf{76.00} & \textbf{121.00} \\
\cline{2-10}
 & & PREEMPT\_RT & No & 4.14 & 4.00 & 51.00 & 1.15 & 4.00 & 6.00 \\
 & & PREEMPT\_RT & \textbf{Yes} & \textbf{41.08} & \textbf{31.00} & \textbf{182.00} & \textbf{27.01} & \textbf{82.00} & \textbf{122.00} \\
\hline
\hline
\textbf{DEADLINE} & R400, D4000 & Standard & No & 6.53 & 7.00 & 33.00 & 2.23 & 8.00 & 16.00 \\
 & & Standard & \textbf{Yes} & \textbf{43.77} & \textbf{36.00} & \textbf{345.00} & \textbf{27.17} & \textbf{78.00} & \textbf{137.00} \\
\cline{2-10}
 & & PREEMPT\_RT & No & 4.20 & 4.00 & 96.00 & 1.86 & 5.00 & 7.00 \\
 & & PREEMPT\_RT & \textbf{Yes} & \textbf{44.16} & \textbf{33.00} & \textbf{209.00} & \textbf{29.57} & \textbf{90.00} & \textbf{137.00} \\
\hline
\textbf{DEADLINE} & R800, D4000 & Standard & No & 6.17 & 6.00 & 58.00 & 1.98 & 7.00 & 14.00 \\
 & & Standard & \textbf{Yes} & \textbf{40.94} & \textbf{35.00} & \textbf{443.00} & \textbf{22.67} & \textbf{70.00} & \textbf{112.00} \\
\cline{2-10}
 & & PREEMPT\_RT & No & 4.21 & 4.00 & 84.00 & 2.00 & 4.00 & 8.00 \\
 & & PREEMPT\_RT & \textbf{Yes} & \textbf{32.30} & \textbf{27.00} & \textbf{197.00} & \textbf{18.75} & \textbf{51.00} & \textbf{116.00} \\
\hline
\end{tabular}%
}
\end{table}

\subsection{Statistical Distribution of Latency}

\begin{figure}[htbp]
    \centering
    \includegraphics[width=\columnwidth]{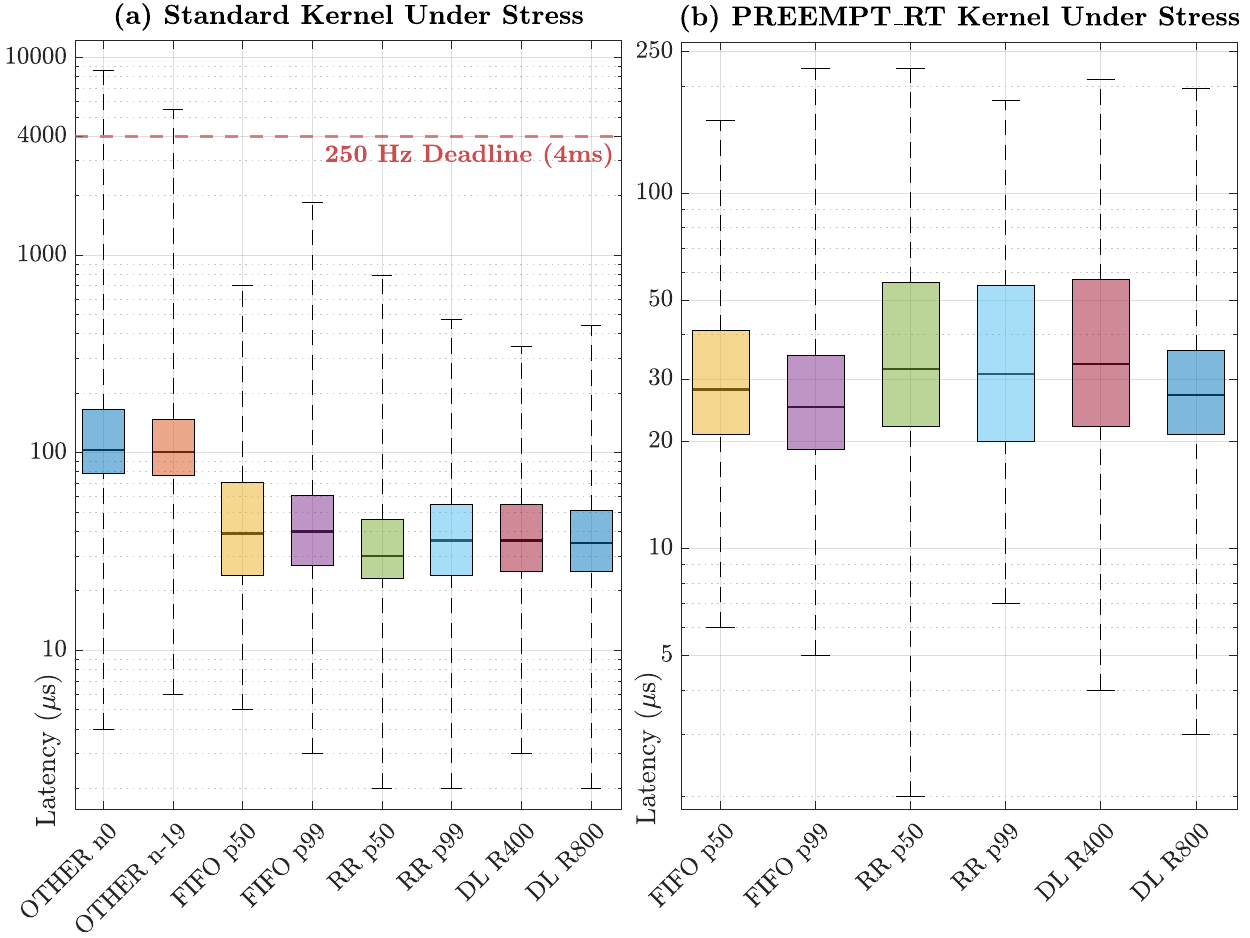}
    \vspace{-5mm}
    \caption{Box plots comparing scheduling latency on the standard and \texttt{PREEMPT\_RT} kernels under heavy system stress. The red dashed line indicates the deadline for the control loop. (a) On the standard kernel, the logarithmic y-axis to visualize the outliers that surpass the deadline. Even real-time policies exhibit significant jitter. (b) On the \texttt{PREEMPT\_RT} kernel, shows the  performance of all real-time schedulers, with worst-case latencies remaining an order of magnitude below the deadline.} 
    \label{fig:rt_schedulers_boxplot}
\end{figure}

Figure \ref{fig:rt_schedulers_boxplot} provides a visual comparison of scheduling latency on both kernels when under heavy system stress.

On the standard kernel, Fig.~\ref{fig:rt_schedulers_boxplot}(a), the default \texttt{SCHED\_OTHER} scheduler exhibits very high worst-case latencies. Table~\ref{tab:scheduling_performance} shows these latencies reach over 9400\,$\mu$s, far exceeding the 4000\,$\mu$s deadline. The real-time schedulers also show significant latency spikes on the standard kernel. For \texttt{SCHED\_FIFO} with priority 99, the maximum observed latency was 1848\,$\mu$s. For \texttt{SCHED\_RR} with priority 99, the maximum was 472\,$\mu$s, and for \texttt{SCHED\_DEADLINE} R800, it was 443\,$\mu$s.

On the \texttt{PREEMPT\_RT} kernel, in Fig.\ref{fig:rt_schedulers_boxplot}(b), the maximum latencies are significantly lower for all real-time policies. The data in Tab.~\ref{tab:scheduling_performance} quantifies this improvement: the maximum latency for \texttt{SCHED\_FIFO} p99 was reduced to 224\,$\mu$s; for \texttt{SCHED\_RR} p99, it was reduced to 182\,$\mu$s; and for \texttt{SCHED\_DEADLINE} R800, it was reduced to 197\,$\mu$s.

\subsection{Iteration-by-Iteration Latency Behavior}

\begin{figure}[h!]
    \centering
    \includegraphics[width=\columnwidth]{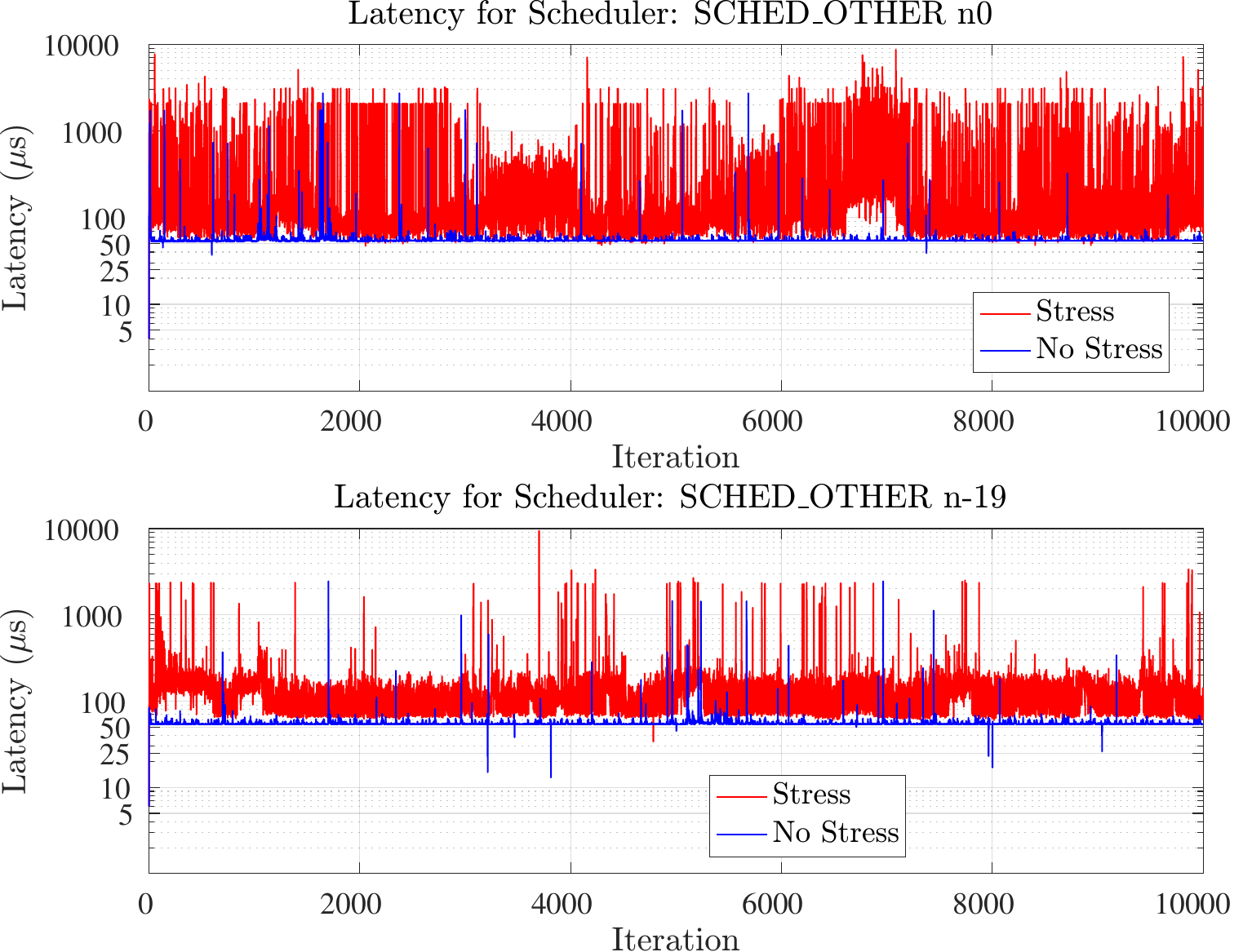}
    \vspace{-6mm}
    \caption{Time-series latency under \texttt{SCHED\_OTHER} on the standard kernel: standard priority (nice 0, top) and high priority (nice -19, bottom). The plots show increased latency and jitter under system stress (red) compared to the idle state (blue).}
    \label{fig:sched_other_latency}
\end{figure}

\begin{figure}[h!]
    \centering
    \includegraphics[width=\columnwidth]{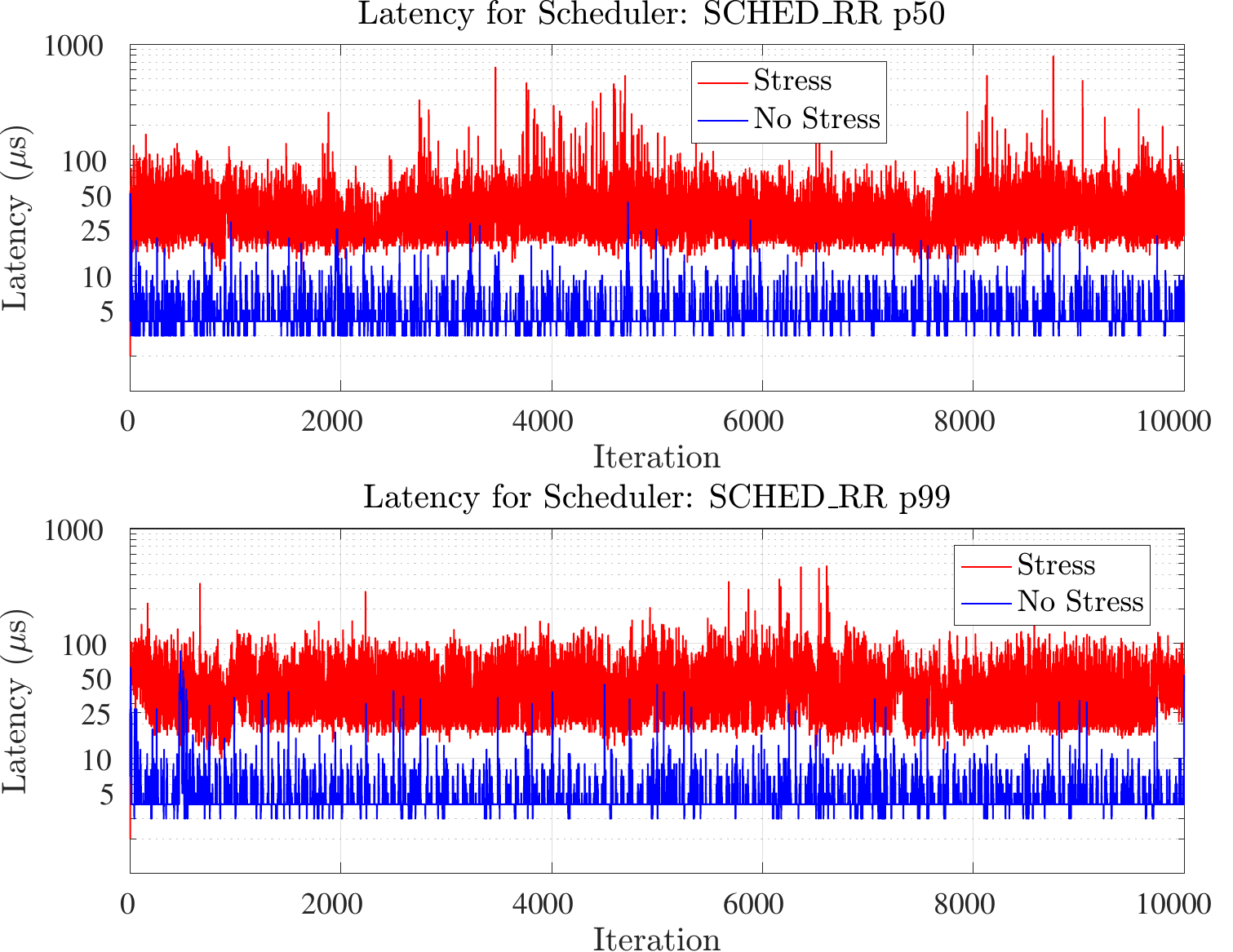}
    \vspace{-6mm}
    \caption{Latency for \texttt{SCHED\_RR} on the standard kernel at priorities 50 (top) and 99 (bottom). Note the spikes under system stress (red), indicating difficulty in guaranteeing deterministic execution even at the highest static priority.}
    \label{fig:sched_rr_latency}
\end{figure}

\begin{figure}[h!]
    \centering
    \includegraphics[width=\columnwidth]{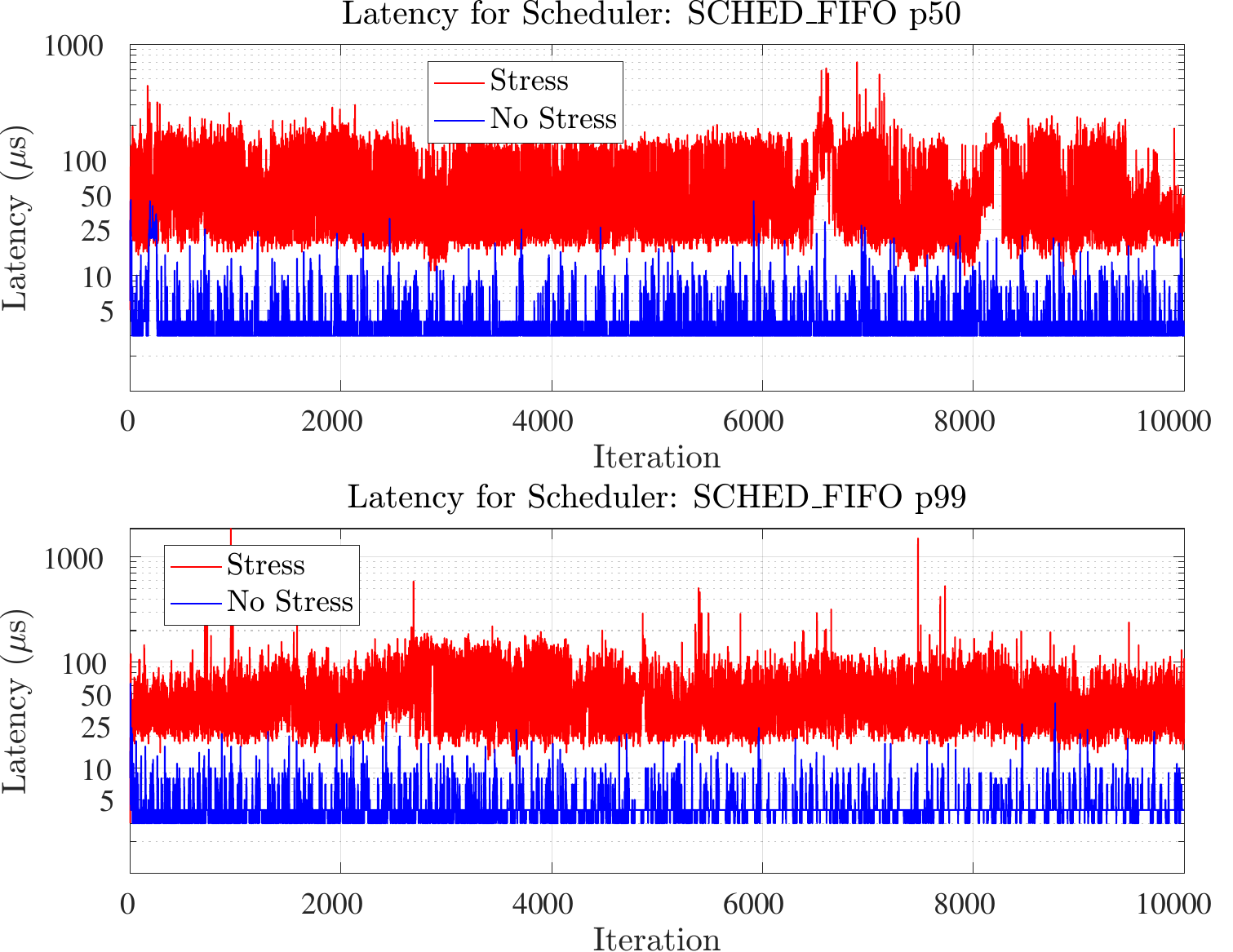}
    \vspace{-6mm}
    \caption{Time-series latency for \texttt{SCHED\_FIFO} on the standard kernel at priorities 50 (top) and 99 (bottom). Similar to \texttt{SCHED\_RR}, the system exhibits latency outliers when under load (red).}
    \label{fig:sched_fifo_latency}
\end{figure}

\begin{figure}[h!]
    \centering
    \includegraphics[width=\columnwidth]{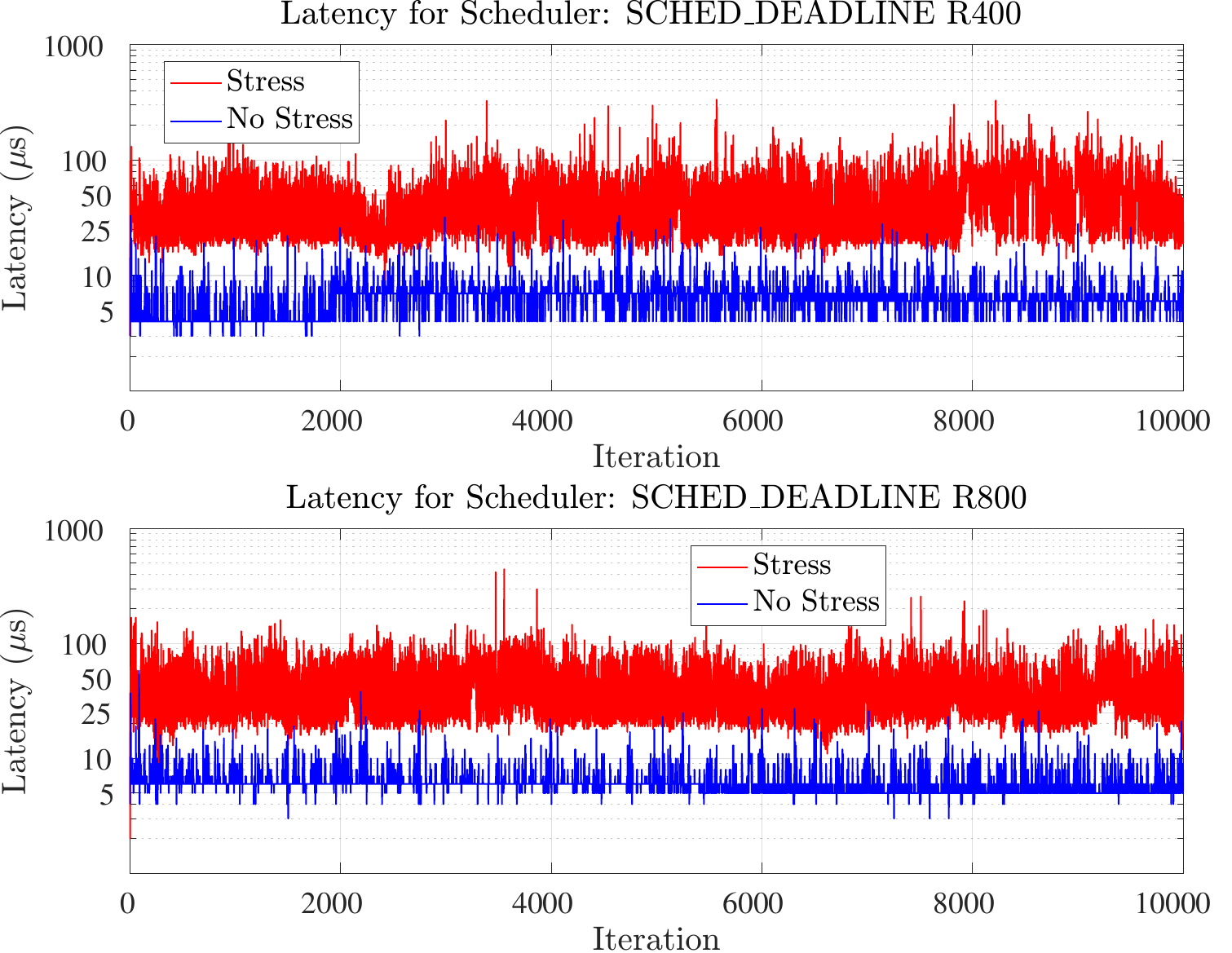}
    \vspace{-6mm}
    \caption{Time-series latency for \texttt{SCHED\_DEADLINE} on the standard kernel, 400\,$\mu$s runtime  (top) versus 800\,$\mu$s (bottom). Even with this reservation-based scheduler, there are significant latency spikes under stress (red).}
    \label{fig:sched_deadline_latency}
\end{figure}

\begin{figure}[h!]
    \centering
    \includegraphics[width=\columnwidth]{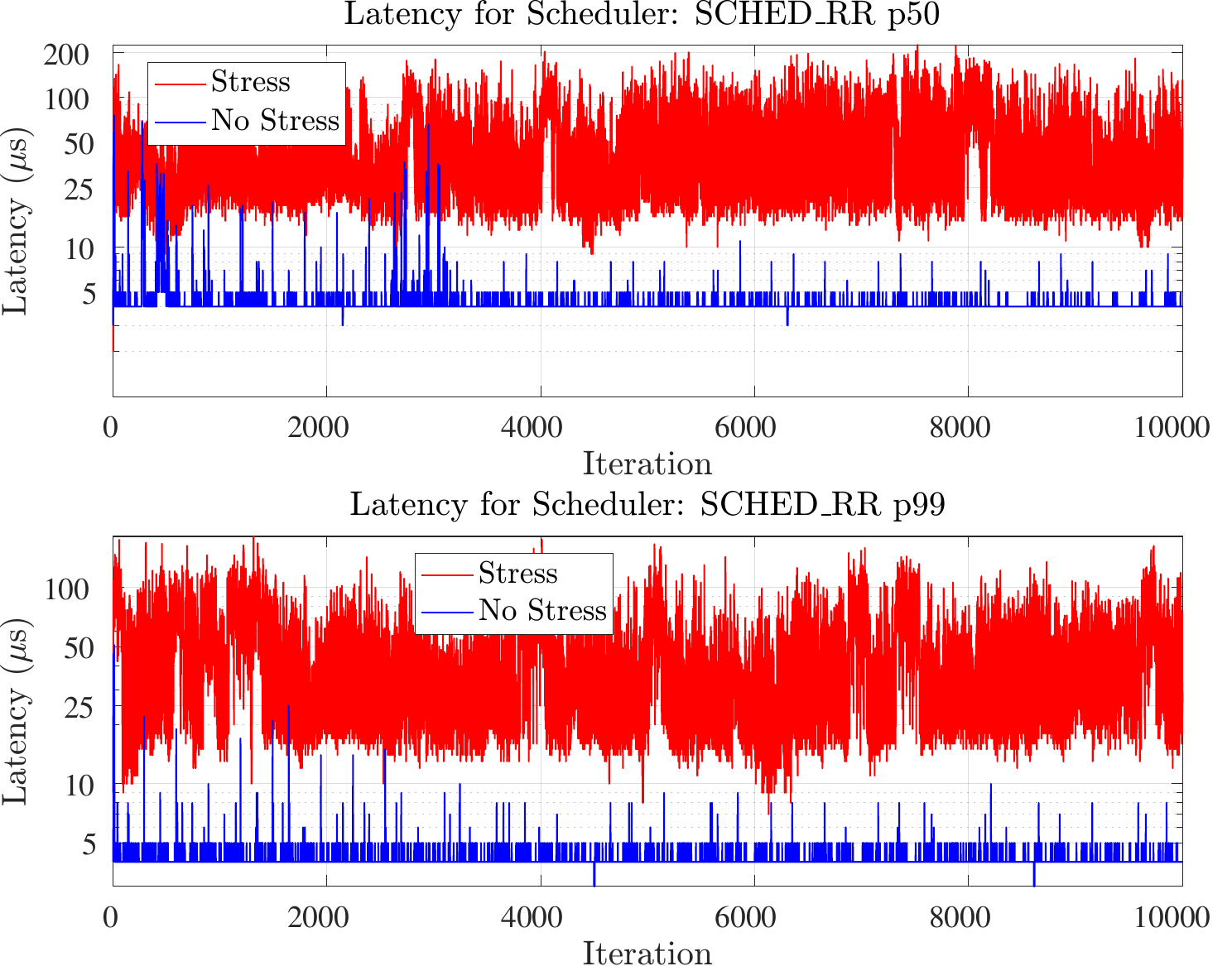}
    \vspace{-6mm}
    \caption{Time-series latency for \texttt{SCHED\_RR} on the \texttt{PREEMPT\_RT} kernel, priority 50 (top) and 99 (bottom). In contrast to the standard kernel, the latency remains tightly bounded with minimal outliers, even when subjected to heavy system stress (red).}
    \label{fig:sched_rr_latency_rt}
\end{figure}

\begin{figure}[h!]
    \centering
    \includegraphics[width=\columnwidth]{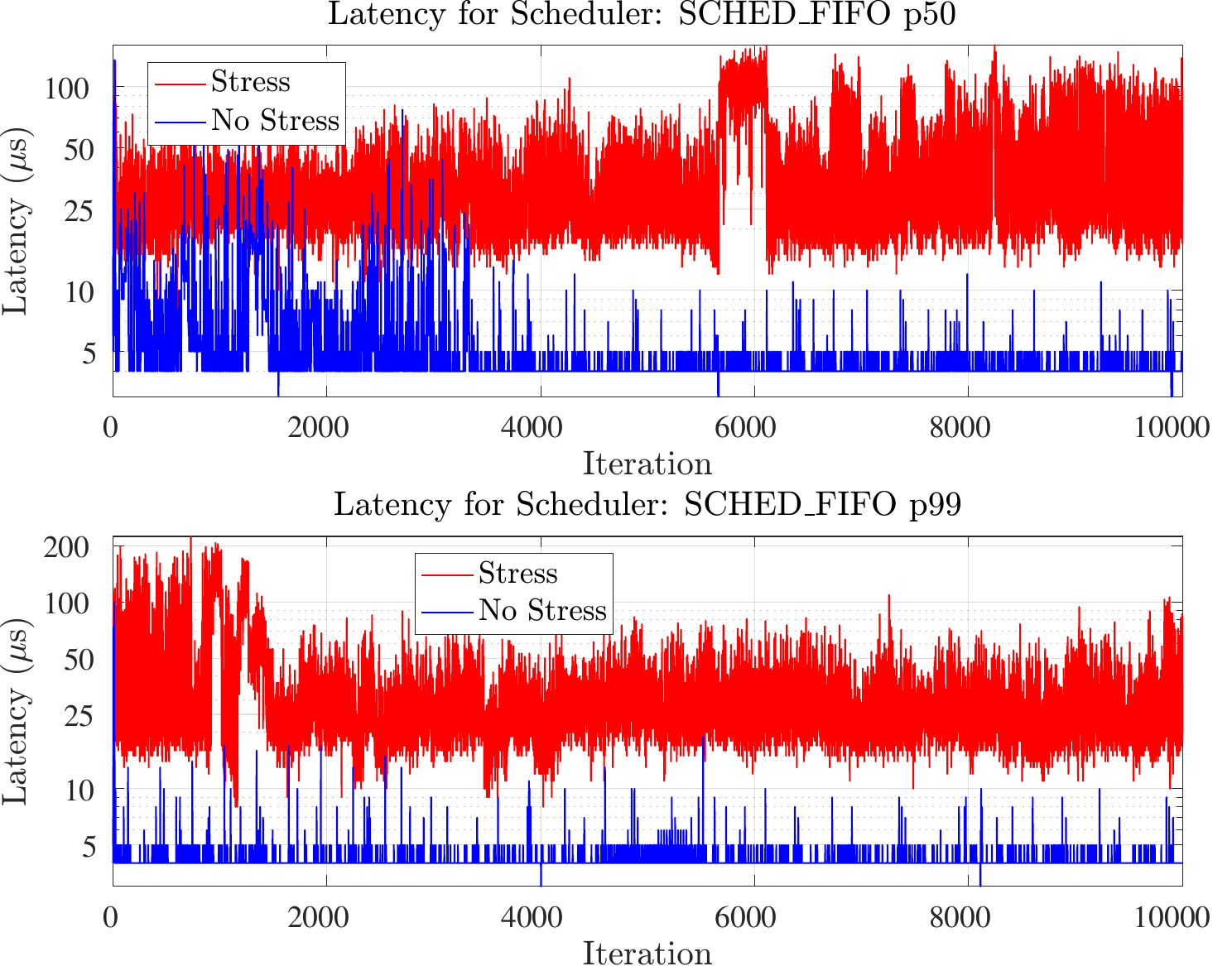}
    \vspace{-6mm}
    \caption{Latency for \texttt{SCHED\_FIFO} on the \texttt{PREEMPT\_RT} kernel for priorities 50 (top) and 99 (bottom). It shows a reduction in the magnitude of outliers under stress (red), demonstrating the improved effectiveness of the real-time kernel's preemption model.}
    \label{fig:sched_fifo_latency_rt}
\end{figure}

\begin{figure}[h!]
    \centering
    \includegraphics[width=\columnwidth]{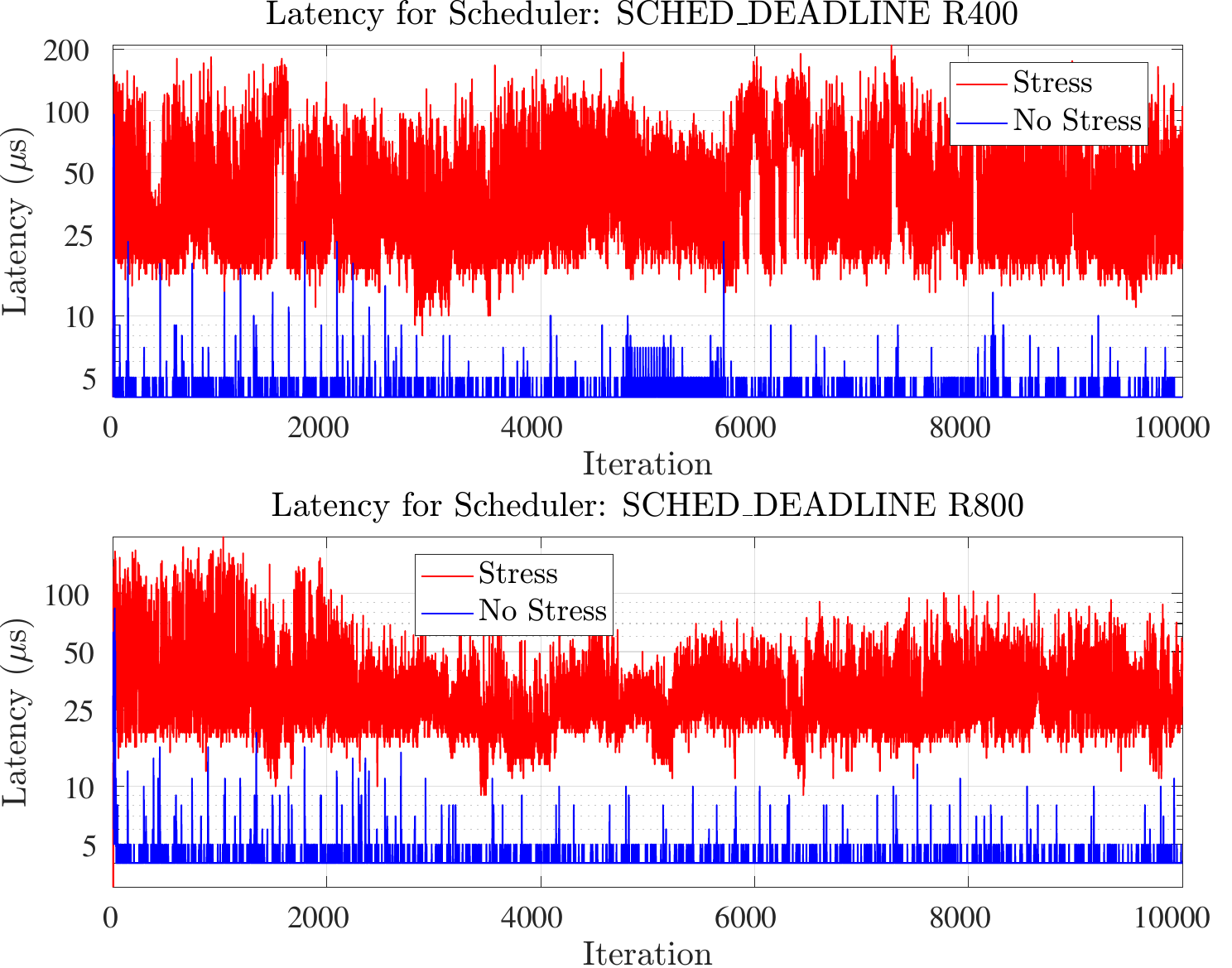}
    \vspace{-6mm}
    \caption{Time-series latency for \texttt{SCHED\_DEADLINE} on the \texttt{PREEMPT\_RT} kernel, comparing a 400\,$\mu$s runtime budget (top) with an 800\,$\mu$s budget (bottom). This policy exhibits the highest stability, with minimal jitter and tightly controlled worst-case latencies under stress (red).}
    \label{fig:sched_deadline_latency_rt}
\end{figure}

To provide a detailed view of system responsiveness, Figs.~\ref{fig:sched_other_latency} through \ref{fig:sched_deadline_latency_rt} depict the latency across 10,000 control loop iterations under different configurations. The standard kernel traces exhibit frequent spikes in execution time, indicating variability and reduced predictability. In contrast, the \texttt{PREEMPT\_RT} kernel traces appear markedly more stable, maintaining a consistent baseline with only occasional small outliers, which are less pronounced in comparison. The standard kernel results are presented in Figs.~\ref{fig:sched_other_latency}-\ref{fig:sched_deadline_latency}, while the corresponding \texttt{PREEMPT\_RT} results are shown in Figs.~\ref{fig:sched_rr_latency_rt}-\ref{fig:sched_deadline_latency_rt}.

\subsection{Analysis of \texttt{SCHED\_OTHER} Non-Determinism}
\label{subsec:root-cause-results}

To investigate the additional latencies with \texttt{SCHED\_OTHER}, the follow-up experiments in Section~\ref{subsec:analysis-methodology} were conducted. The first test isolated the impact of system-level services. As shown in Tab.~\ref{tab:gui-impact}, disabling the GUI reduced the worst-case latency for \texttt{SCHED\_OTHER} under stress by up to 61.4\%. Despite this improvement, the maximum latency for the high-priority task (nice -19) still reached 3635\,$\mu$s. This demonstrates that while high-level services are a significant source of interference, they do not account for all of the observed non-determinism.

\begin{table}[htpb]
\centering
\caption{Impact of Disabling the Graphical User Interface (GUI) on \texttt{SCHED\_OTHER} Worst-Case Latency Under System Stress.}
\label{tab:gui-impact}
\resizebox{\columnwidth}{!}{%
\begin{tabular}{@{}l|cc|c@{}}
\toprule
\textbf{Scheduler (Config)} & \textbf{Max Latency (GUI Enabled)} & \textbf{Max Latency (GUI Disabled)} & \textbf{Improvement (\%)} \\ \midrule
\texttt{SCHED\_OTHER} (Nice -19) & 9424\,$\mu$s & \textbf{3635\,$\mu$s} & \textbf{-61.4\%} \\
\texttt{SCHED\_OTHER} (Nice 0)   & 8626\,$\mu$s & \textbf{5993\,$\mu$s} & \textbf{-30.5\%} \\ \bottomrule
\end{tabular}%
}
\end{table}

To identify the cause of the remaining latency, we utilized \texttt{perf} to trace the exact activation sequence of the control task. Tab.~\ref{tab:trace_sequence} contrasts the timestamped sequence of events captured during a single control loop activation.

\begin{table}[htpb]
\centering
\caption{Kernel Activation Sequence  (Trace Data). Timestamps denote $\Delta t$ from the physical interrupt arrival. The "Deferred Path" in the standard kernel introduces significant delays ($>100\,\mu$s) even before the task begins execution.}
\label{tab:trace_sequence}
\resizebox{\columnwidth}{!}{%
\begin{tabular}{@{}ll|ll@{}}
\toprule
\multicolumn{2}{c|}{\textbf{PREEMPT\_RT (Direct Path)}} & \multicolumn{2}{c}{\textbf{Standard Kernel (Deferred Path)}} \\ \midrule
\textbf{$\Delta t$ ($\mu$s)} & \textbf{Event} & \textbf{$\Delta t$ ($\mu$s)} & \textbf{Event} \\ \midrule
0 & \texttt{irq\_handler\_entry} & 0 & \texttt{irq\_handler\_entry} \\
2 & \texttt{sched\_wakeup} (target: \textbf{fcs}) & 5 & \texttt{irq\_softirq\_raise} (HRTIMER) \\
4 & \texttt{sched\_switch} ($\to$ \textbf{fcs}) & 17 & \texttt{sched\_wakeup} (target: \textbf{ktimers}) \\
\textbf{7} & \textbf{Task Execution Begins} & 54 & \texttt{sched\_switch} ($\to$ \textbf{ktimers}) \\
- & - & 92 & \texttt{timer\_expire} (ktimers) \\
- & - & 93 & \texttt{sched\_wakeup} (target: \textbf{fcs}) \\
- & - & \textbf{117} & \textbf{Task Execution Begins} \\ \bottomrule
\end{tabular}%
}
\end{table}

The traces reveal a contrast in the wake-up mechanism. The \texttt{PREEMPT\_RT} kernel allows the hardware timer interrupt to wake the user task directly, resulting in a wake-up latency of just 7\,$\mu$s. In contrast, the standard kernel employs a \textit{Deferred Path}: the interrupt offloads the timer expiry to the intermediate \texttt{ktimers} thread (softirq). In the trace shown, this indirection introduced a delay of over 100\,$\mu$s and required three additional context switches before the flight control task was scheduled.

The traces also revealed a secondary effect: the computational execution time of the control algorithm itself (excluding scheduling delay) increased from $\approx 3.8\,\mu$s in the real-time kernel to $\approx 51\,\mu$s in the standard kernel. Since the code payload is identical, this $13\times$ slowdown suggests cache pollution. The intermediate execution of kernel housekeeping threads (`ktimers`, `kworker`) flushes instructions and data from the L1/L2 caches, forcing the control task to fetch data from slower main memory when it finally runs. This indicates that the standard kernel suffers not only from scheduling latency but also from induced memory hierarchy contention.

\section{Discussion and Architectural Analysis}
\label{sec:Discussion}

The data reveals a distinct separation between the kernel's ability to schedule tasks (software) and the platform's ability to serve memory requests (hardware). This section interprets the results through an architectural lens, isolating the sources of determinism and establishing a feasibility baseline relative to traditional microcontroller-based architectures.

\subsection{Mechanism of Determinism: Direct vs. Deferred Activation}
\label{subsec:architectural-source}
The primary source of the $>9$\,ms latency spikes observed in the standard kernel (Tab. \ref{tab:scheduling_performance}) is the Deferred Activation Path. In standard Linux, a timer interrupt does not immediately wake the user-space task but schedules a Software Interrupt (SoftIRQ), often deferred to \texttt{ksoftirqd} to maintain throughput. Under stress, runqueue lock contention delays this SoftIRQ, decoupling the physical interrupt from the task wake-up.

In contrast, \texttt{PREEMPT\_RT} enforces a Direct Activation Path. By converting interrupt handlers into preemptible threads, the patch flattens the activation hierarchy. Our traces (Table VI) show that the critical path becomes $IRQ \rightarrow Task$, bypassing the non-deterministic SoftIRQ layer. This architectural change drives the 87.9\% reduction in worst-case latency for \texttt{SCHED\_FIFO}, ensuring the sub-millisecond response required for 250\,Hz control.

\subsection{The Residual Jitter: Memory Hierarchy Contention}
While \texttt{PREEMPT\_RT} bounds OS scheduling latency, a residual worst-case jitter of 200--225\,$\mu$s persists even on isolated CPU cores. This points to the architectural complexity of the Pi 5's BCM2712 SoC, rather than the kernel scheduler.

Unlike MCUs with dedicated SRAM, the quad-core Cortex-A76 shares a unified L3 cache and DRAM controller. We hypothesize that under the synthetic memory stress profile, "noisy neighbor" threads on Cores 0--1 induce L3 cache thrashing and memory bus contention. This mechanism is supported by the $13\times$ execution time dilation of the control payload itself (from $\approx 3.8\,\mu$s to $\approx 51\,\mu$s) observed in the trace analysis. Even with the flight control task pinned to Core 2, it likely suffers memory stall cycles, forced to fetch evicted instructions and data from slower main memory. Consequently, on modern edge-AI SoCs, reducing jitter to MCU-levels ($<50\,\mu$s) requires more than just kernel preemption; it needs hardware-level spatial partitioning (e.g., cache coloring), which remains a complex and open challenge in Linux.

\subsection{Feasibility Analysis: RT-Linux vs. MCU Baseline}
To contextualize these results for UAV designers, we compare the measured performance of the Raspberry Pi 5 against the industry-standard microcontroller (MCU) baseline, typified by an STM32F7 running an RTOS (e.g., FreeRTOS/NuttX).

\begin{table}[h]
\centering
\caption{Architectural Comparison: RPi 5 (Measured) vs. Typical MCU Flight Controller. RPi 5 values are from our experimental data; STM32 values are reference metrics for Cortex-M7 RTOS implementations from literature \cite{9999639, PX4_Arch}.}
\label{tab:rtos_comparison}
\resizebox{0.85\columnwidth}{!}{%
\begin{tabular}{|l|c|c|}
\hline
\textbf{Metric} & \textbf{RPi 5 (RT-Linux)} & \textbf{STM32F7 (RTOS) \cite{9999639, PX4_Arch}} \\
\hline
\hline
\textbf{Clock Speed} & 2.4 GHz & 216 MHz \\
\textbf{Worst-Case Jitter} & $\approx 224\,\mu s$ & $\approx 10-20\,\mu s$ \\
\textbf{Control Period (250Hz)} & 4000\,$\mu s$ & 4000\,$\mu s$ \\
\textbf{Jitter \% of Period} & \textbf{5.6\%} & \textbf{0.5\%} \\
\textbf{Compute Headroom} & High (Vision/AI) & Low (Control Only) \\
\hline
\end{tabular}%
}
\end{table}

As seen in Table \ref{tab:rtos_comparison}, the \texttt{PREEMPT\_RT} Linux solution exhibits approximately $10\times$ higher jitter than a bare-metal RTOS \cite{PX4_Arch}. However, in the context of a 250\,Hz control loop ($T=4000\,\mu s$), the $224\,\mu s$ worst-case jitter represents only 5.6\% of the timing budget. For flight control stability, this margin is generally acceptable \cite{control_design_reference}. 
This confirms that while Linux cannot match the microsecond-level precision of an MCU, the Raspberry Pi 5 with \texttt{PREEMPT\_RT} offers \textit{sufficient} determinism for the control loop while unlocking orders of magnitude more computational power for integrated autonomy.

\subsection{Implications for Flight Stability}
Flight stability is contingent on the controller's ability to react to disturbances within a bounded phase margin. The standard kernel exhibited worst-case latencies of $9.4$\,ms, exceeding the $4.0$\,ms period of a 250\,Hz attitude loop by a factor of $2.3\times$. In a physical system, this control blackout forces actuators to hold outdated commands for multiple cycles, effectively opening the control loop. During aggressive maneuvers, such delays lead to state divergence and potential loss of control.
Conversely, the \texttt{PREEMPT\_RT} kernel bounds the worst-case jitter to $225\,\mu$s ($5.6\%$ of the period). This ensures that the control law is computed with fresh state estimates in every cycle, maintaining the phase margin required for safe flight even under heavy CPU saturation.

\subsection{I/O Latency Considerations}
\label{subsec:io_considerations}
While this study establishes that \texttt{PREEMPT\_RT} bounds \textit{computational} scheduling latency to levels compatible with 250\,Hz control ($<225\,\mu$s), scheduler determinism is a necessary but insufficient condition for flight stability. End-to-end performance remains dependent on the I/O subsystem. In unified architectures, standard Linux device drivers can introduce unbounded latencies via spinlocks or non-threaded interrupts, potentially negating the scheduler's guaranties. Consequently, achieving the theoretical performance demonstrated in this work likely requires bypassing standard kernel driver stacks in favor of Userspace I/O or DMA-based transfer mechanisms. This isolates the high-frequency control loop from the blocking behavior inherent to standard Linux peripherals.

\section{Conclusion}
\label{sec:Conclusion}

Our work evaluated the scheduling performance of the \texttt{PREEMPT\_RT} Linux kernel against the standard kernel for a 250\,Hz UAV flight control task on a Raspberry Pi 5. The results show that the standard kernel is unsuitable for this role. Its default scheduler produced worst-case latencies above 8\,ms, and its POSIX real-time schedulers exhibited latency spikes up to 1848\,$\mu$s under stress, limiting their usage for safety-critical control. 

In contrast, the \texttt{PREEMPT\_RT} kernel, when paired with any of the real-time policies (\texttt{SCHED\_FIFO}, \texttt{SCHED\_RR}, or \texttt{SCHED\_DEADLINE}), consistently bounded worst-case latencies at or below 225\,$\mu$s, even under heavy load. These findings demonstrate that a properly configured GPOS with \texttt{PREEMPT\_RT} provides the timing predictability required for high-frequency control. Moreover, because this behavior is tied to the threaded-IRQ mechanism introduced by the patch, these results suggest similar behavior in recent real-time Linux kernel versions.

The results suggest that unified GPOS architectures on a single board are a viable alternative to dual-processor designs in many UAV applications, especially where constraints are less stringent. However, this study intentionally isolated CPU scheduling from I/O. Future work should integrate real-world sensor and actuator I/O to evaluate the combined effects of scheduler latency and hardware interrupt handling. The analysis should also extend to more computationally demanding controllers, such as Model Predictive Control (MPC) and Sliding Mode Control (SMC). Finally, the trade-off between real-time performance and power consumption should be explored, as achieving low latency often requires disabling power-saving features essential for battery-powered platforms.

\bibliographystyle{unsrt}  
\bibliography{references}  

@article{adam2021real,
  title={Real-time performance and response latency measurements of Linux kernels on single-board computers},
  author={Adam, George K},
  journal={Computers},
  volume={10},
  number={5},
  pages={64},
  year={2021},
  publisher={MDPI}
}

@inproceedings{iorga2020slow,
  title={Slow and steady: Measuring and tuning multicore interference},
  author={Iorga, Dan and Sorensen, Tyler and Wickerson, John and Donaldson, Alastair F},
  booktitle={2020 IEEE Real-Time and Embedded Technology and Applications Symposium (RTAS)},
  pages={1--13},
  year={2020},
  organization={IEEE}
}

@inproceedings{yang2020obtaining,
  title={Obtaining hard real-time performance and rich Linux features in a compounded real-time operating system by a partitioning hypervisor},
  author={Yang, Chung-Fan and Shinjo, Yasushi},
  booktitle={Proceedings of the 16th ACM SIGPLAN/SIGOPS International Conference on Virtual Execution Environments},
  pages={59--72},
  year={2020}
}

@ARTICLE{CETDS_IEEE,
  author={Ricardo, Jorge A. and Giacomossi, Luiz and Trentin, João F. S. and Brancalion, José F. B. and Maximo, Marcos R. O. A. and Santos, Davi A.},
  journal={IEEE Access}, 
  title={Cooperative Threat Engagement Using Drone Swarms}, 
  year={2023},
  volume={11},
  number={},
  pages={9529-9546},
  doi={10.1109/ACCESS.2023.3239817}}

@phdthesis{bouabdallah2007design,
  author  = {Bouabdallah, Samir},
  title   = {Design and control of quadrotors with application to autonomous flying},
  school  = {\'Ecole Polytechnique F\'ed\'erale de Lausanne},
  year    = {2007},
  doi     = {10.5075/epfl-thesis-3727}
}

@InProceedings{Search_IAI,
author="Giacomossi, Luiz
and Maximo, Marcos R. O. A.
and Sundelius, Nils
and Funk, Peter
and Brancalion, Jos{\'e} F. B.
and Sohlberg, Rickard",
title="Cooperative Search and Rescue with Drone Swarm",
booktitle="International Congress and Workshop on Industrial AI and eMaintenance 2023",
year="2024",
publisher="Springer Nature Switzerland",
address="Cham",
pages="381--393",
isbn="978-3-031-39619-9"
}

@bachelorsthesis{martensson2016flying,
  title        = {Flying Penguins: Building and Evaluating the Viability of a Linux-based Drone},
  author       = {Mårtensson, Anders},
  school       = {Blekinge Institute of Technology, Faculty of Computing, Department of Computer Science and Engineering},
  year         = {2016},
  type         = {Bachelor's thesis},
  note         = {Bachelor Thesis in Computer Science, Educational program: DVGDS Computer and System Science},
  supervisor   = {Lopez Luro, Francisco}
}

@INPROCEEDINGS{nvidia_pixhawk,
  author={Qu, Xuanyao and Wei, Ying and Liu, Yonghan and Su, Xuanguang},
  booktitle={2024 Int. Conf. Electr. Drives Power Electron. Eng.}, 
  title={Design of Automatic Search and Rescue UAV Based on Jetson Nano Combined with PX4 Pixhawk Flight Controller and Color Recognition Technology}, 
  year={2024},
  volume={},
  number={},
  pages={460–466},
  doi={10.1109/EDPEE61724.2024.00092}}

@INPROCEEDINGS{RT_Helicopter,
  author={Won Eui Hong and Jae Shin Lee and Rai, L. and Soon Ju Kang},
  booktitle={11th IEEE International Conference on Embedded and Real-Time Computing Systems and Applications (RTCSA'05)}, 
  title={RT-Linux based hard real-time software architecture for unmanned autonomous helicopters}, 
  year={2005},
  volume={},
  number={},
  pages={555-558},
  doi={10.1109/RTCSA.2005.83}}

@inproceedings{brown2010fast,
  title={How fast is fast enough? Choosing between Xenomai and Linux for real-time applications},
  author={Brown, Jeremy H and Martin, Brad},
  booktitle={Proceedings of the 12th Real-Time Linux Workshop (RTLWS)},
  pages={1--17},
  year={2010}
}

@INPROCEEDINGS{PX4_Arch,
  author={Meier, Lorenz and Honegger, Dominik and Pollefeys, Marc},
  booktitle={2015 IEEE International Conference on Robotics and Automation (ICRA)}, 
  title={PX4: A node-based multithreaded open source robotics framework for deeply embedded platforms}, 
  year={2015},
  volume={},
  number={},
  pages={6235-6240},
  doi={10.1109/ICRA.2015.7140074}}

@article{kangunde2021review,
  title={A review on drones controlled in real-time},
  author={Kangunde, Vemema and Jamisola Jr, Rodrigo S and Theophilus, Emmanuel K},
  journal={International journal of dynamics and control},
  volume={9},
  number={4},
  pages={1832--1846},
  year={2021},
  publisher={Springer}
}

@inproceedings{giernacki2017crazyflie,
  title={Crazyflie 2.0 quadrotor as a platform for research and education in robotics and control engineering},
  author={Giernacki, Wojciech and Skwierczy{\'n}ski, Mateusz and Witwicki, Wojciech and Wro{\'n}ski, Pawe{\l} and Kozierski, Piotr},
  booktitle={2017 22nd international conference on methods and models in automation and robotics (MMAR)},
  pages={37--42},
  year={2017},
  organization={IEEE}
}

@INPROCEEDINGS{linux_raspberry,
  author={Carvalho, Alan and Machado, Cláudio and Moraes, Fabiano},
  booktitle={2019 Latin American Robotics Symposium (LARS)}, 
  title={Raspberry Pi Performance Analysis in Real-Time Applications with the RT-Preempt Patch}, 
  year={2019},
  volume={},
  number={},
  pages={162-167},
  doi={10.1109/LARS-SBR-WRE48964.2019.00036}}

@article{control_design_reference,
title = {Smooth second-order sliding mode control for fully actuated multirotor aerial vehicles},
journal = {ISA Transactions},
volume = {129},
pages = {169-178},
year = {2022},
issn = {0019-0578},
author = {Jorge A. {Ricardo Jr} and Davi A. Santos}
}

@ARTICLE{9999639,
  author={Rico, Ramón and Rico-Azagra, Javier and Gil-Martínez, Montserrat},
  journal={IEEE Access}, 
  title={Hardware and RTOS Design of a Flight Controller for Professional Applications}, 
  year={2022},
  volume={10},
  number={},
  pages={134870-134883},
  doi={10.1109/ACCESS.2022.3232749}}

@article{smeds2012effect,
  author    = {Kristofer Smeds and Xiaodong Lu},
  title     = {Effect of sampling jitter and control jitter on positioning error in motion control systems},
  journal   = {Precision Engineering},
  volume    = {36},
  number    = {2},
  pages     = {175--192},
  year      = {2012},
  issn      = {0141-6359},
  doi       = {10.1016/j.precisioneng.2011.09.002}
}

@INPROCEEDINGS{optimization_fcs,
  author={Giacomossi, Luiz and Caregnato-Neto, Angelo and Maximo, Marcos R. O. A.},
  booktitle={2022 Latin American Robotics Symposium (LARS)}, 
  title={Optimization of Force and Torque Bounds for the Flight Control System of a Quadcopter using PSO}, 
  year={2022},
  volume={},
  number={},
  pages={1-6},
  doi={10.1109/LARS/SBR/WRE56824.2022.9995998}
}

@ARTICLE{survey_uavs,
  author={Javed, Sadaf and Hassan, Ali and Ahmad, Rizwan and Ahmed, Waqas and Ahmed, Rehan and Saadat, Ahsan and Guizani, Mohsen},
  journal={IEEE Internet of Things Journal}, 
  title={State-of-the-Art and Future Research Challenges in UAV Swarms}, 
  year={2024},
  volume={11},
  number={11},
  doi={10.1109/JIOT.2024.3364230}}

@book{beard2012small,
  title={Small unmanned aircraft: Theory and practice},
  author={Beard, Randal W and McLain, Timothy W},
  year={2012},
  publisher={Princeton university press}
}

@article{preempt_survey,
author = {Reghenzani, Federico and Massari, Giuseppe and Fornaciari, William},
title = {The Real-Time Linux Kernel: A Survey on PREEMPT\_RT},
year = {2019},
issue_date = {January 2020},
publisher = {Association for Computing Machinery},
address = {New York, NY, USA},
volume = {52},
number = {1},
issn = {0360-0300},
doi = {10.1145/3297714},
journal = {ACM Comput. Surv.},
month = feb,
articleno = {18},
numpages = {36}
}

@book{love2010linux,
  title={Linux kernel development},
  author={Love, Robert},
  year={2010},
  publisher={Pearson Education}
}

\end{document}